\documentclass[aps,pre,twocolumn,showpacs,longbibliography,floatfix]{revtex4-1}
\usepackage{graphicx}
\usepackage{amssymb}
\usepackage{dcolumn}
\usepackage{bm}
\usepackage{dcolumn}

\def\s#1{_{\rm #1} }

\def\lz{\ell_{\parallel}}
\def\lp{\ell_{\bot}}
\def\Pp{P_{\parallel}}
\def\Pn{P_{\bot}}
\def\bea{\begin{eqnarray}}
\def\eea{\end{eqnarray}}
\def \be{\begin{equation}}
\def \ee{\end{equation}}
\def\nl{\hfil\break}

\def\d{\textrm{d}}
\usepackage{graphicx,xcolor}

\begin{document}
\title{Optomechanical conversion by mechanical turbines}
\author{Milo\v{s} Kne\v{z}evi\'{c}}
\email{mk684@cam.ac.uk}
\author{Mark Warner}
\affiliation{Cavendish Laboratory, University of Cambridge, Cambridge CB3 0HE, United Kingdom}
\date{\today}

\begin{abstract}
Liquid crystal elastomers are rubbers with liquid crystal order. They contract along
their nematic director when heated or illuminated. The shape changes are large and occur in a relatively
narrow temperature interval, or at low illumination, around the nematic--isotropic transition. We present a conceptual design of
a mechanical, turbine-based engine using photo-active liquid crystal elastomers to extract mechanical work from light. Its efficiency
is estimated to be 40\%.
\end{abstract}


\maketitle

\section{Introduction} \label{intro}

\begin{figure}[!b]
  \includegraphics[width=7.7cm]{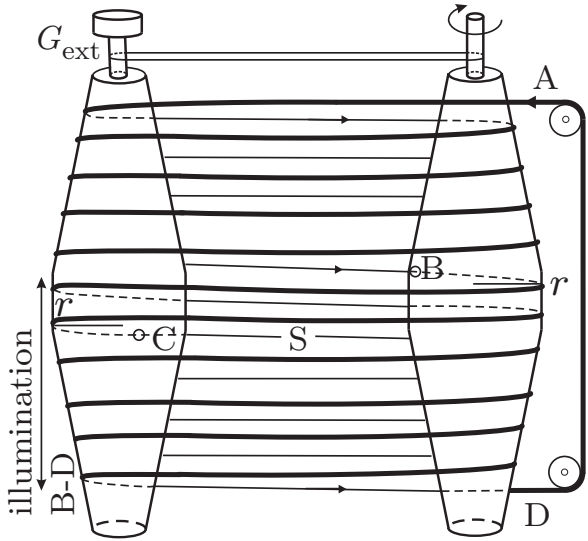}
  \caption{Schematic of an optomechanical turbine.
  }
\label{fig1}
\end{figure}

We propose a mechanical turbine-based engine to harness the contractions of soft, photo-responsive solids with a large stroke.  We thus take photo-active nematic liquid crystal elastomers (LCEs) as our working material. Related engines have been proposed before, for instance based on the bend response of strips of nematic photo-glasses connecting two wheels~\cite{ikeda:08,palffy:13}.
In contrast, we proposed a two-wheel stretch engine~\cite{knezevic:13o} using a nematic photo-LCE.  The engine had analogies to that of Steinberg \textit{et. al} who studied chemical to mechanical energy conversion~\cite{steinberg:66}.  Here, we extend the two-wheel approach of~\cite{knezevic:13o} to mechanical turbine-based engines, see Fig.~\ref{fig1}, in order to get a much higher conversion of light to mechanical work than in that paper.
There is a similarity to a turbine that converted chemical to mechanical
energy~\cite{sussman:70}.  Here we analyse the mechanics and losses involved in such turbines.
Modelling the geometrical and material parameters of this turbine-based engine, along with the known photo-response of typical LCEs, suggests that its efficiency can be as high as $40\%$.

The basis of these two-wheel and turbine engines is that a nematic rubber strip passes respectively once around two wheels, or multiply around spindles, turning at the same rate. The strip reduces its natural length on heating or illumination during its transition around one wheel; since its length while on the wheel is fixed, then the tension rises.  On leaving the wheel, it physically contracts to its new equilibrium length and, in thereby eliminating the tension that has arisen, does work on the wheels/spindles that are turning underneath it.  A net work is done because the wheels are of different diameters or the spindles are tapered.  This paper is about the mechanics, work cycle and efficiency if spindles are used rather than wheels.  With many passes around the spindles, the work cycle comes to resemble that of a turbine.

Classical elastomers are cross-linked polymer melts exhibiting liquid characteristics locally,
but are solid-like on a macroscopic scale. Incorporating molecular rods into the polymers of a simple elastomer leads to
networks that combine orientational liquid crystal order with the extreme stretchiness of rubber, that is LCEs~\cite{deGennes:75,warnerbook:07}.
The shape of a monodomain LCE is very sensitive to the change of the nematic order parameter $Q$; network polymers are elongated by the directional order and mechanical shape change ensues.
The order decreases on increasing the temperature, which manifests as a uniaxial contraction, by a factor $\lambda\s{m}$ ($<1$) of the
elastomer along the nematic director~\cite{warnerbook:07}. The contraction is especially
rapid in the vicinity of the transition temperature to the isotropic state.
Analogous shape changes occur in photoelastomers in which photoisomerizable dye molecules, rod-like in their \textit{trans} ground state, are connected to the LCE structure~\cite{finkelmann:01,hogan:02}.
Here, illumination causes the creation of bent-shaped \textit{cis} isomers of the dye which act as impurities
that reduce the nematic order, in turn leading to a contraction of the photoelastomer. The presence of \textit{cis}
isomers raises the effective temperature of the photoelastomer from $T$ to a pseudo $\tilde{T} > T$ which depends on the \textit{cis} concentration and mimics the disorder as if it were induced thermally~\cite{finkelmann:01,knezevic:13}. It is important to note
that mechanical deformations of elastomers are reversible, that is, on removal of heat or light, recovery elongations by a factor of $1/\lambda\s{m}$ occur. These elongations can be huge, up to 400\%~\cite{tajbakhsh:01}.

\section{Operating principle} \label{oprinciple}

\begin{figure}[!t]
  \includegraphics[width=8.7cm]{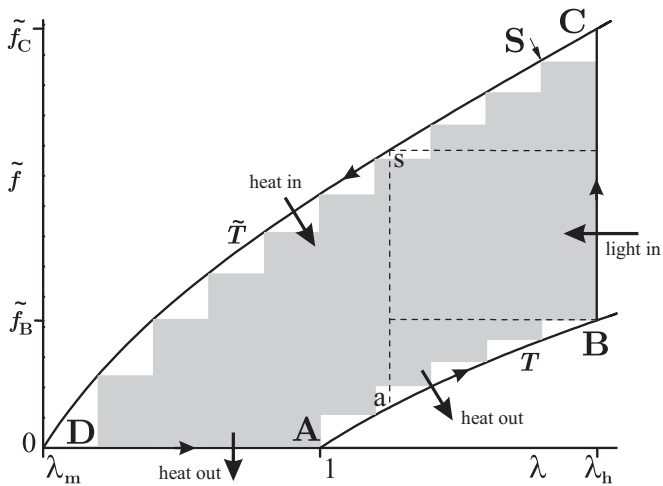}
  \caption{The optothermal cycle of the turbine-based engine. Reduced force $\tilde{f}$ against stretch $\lambda$. Upper curve --- pseudo temperature $\tilde{T}$ (illuminated, isotropic state), bottom --- lower $T$ (nematic state).
  The grayed area represents the work done by the turbine-based engine having a finite number of extension and contraction steps (to be discussed in Section \ref{friction}). Dashed lines --- two-wheel motor~\cite{knezevic:13o}.
  }
\label{fig2}
\end{figure}

The optothermal cycle of our engine is shown in Fig.~\ref{fig2}.
For a nematic elastomer of shear modulus $\mu$, the free energy per length of unstretched band, $F$, and the tension $f$ depend on the stretch ratio (deformation gradient) $\lambda$~\cite{warnerbook:07}:
\bea
F(\lambda,T) &=& \frac{1}{2} \mu A_0 \left ( \Pp \lambda^2 + \frac{2 \Pn}{\lambda} \right ), \\
\tilde{f} &=& \frac{f}{ \mu A_0} =  \frac{1}{\mu A_0}\left ( \frac{\partial F}{\partial \lambda} \right )_T =  \left ( \Pp \lambda - \frac{\Pn}{\lambda^2} \right ),
\label{nrubber}
\eea
where $A_0$ is the cross sectional area of the unstretched elastomer, and $\tilde{f} = f/(\mu A_0)$ is the tension reduced by the natural force scale in the problem. The modulus $\mu$ is temperature dependent.
A simple, freely-jointed rod (FJR) model quite accurately describes a wide range of LCEs~\cite{finkelmann:01t,warnerbook:07} and, in particular, the development of photoforce~\cite{knezevic:13}.
In this model, the coefficients $\Pp = \lz/\tilde{\ell}_{\parallel}$ and $\Pn=\lp/\tilde{\ell}_{\bot}$ are the ratios of the effective step lengths of the network polymers at the reference temperature $T$ and the current temperature $\tilde{T}$, respectively parallel and perpendicular to the director.  Before heating or illumination, the temperature is $T$; the extension factors $\lambda$ are measured from the lengths in this initial state.    The step lengths depend directly on the aspect ratio of the units of the polymer, but these cancel in the ratios $P$, thus thus depend on the nematic order parameters $Q(T)$ and $\tilde{Q} = Q(\tilde{T})$ in a simple way in the FJR model:
\be \Pp = (1+2Q)/(1+2\tilde{Q})\;\;\textrm{and}
\;\; \Pn = (1-Q)/(1-\tilde{Q})\label{P-defs}.
\ee Note that the aspect ratio of the polymer units also contributes to the order $Q$.
Along A$\rightarrow$B, the current temperature is actually $\tilde{T} = T$ (nothing has changed) and $\Pp = 1$ and $\Pn= 1$ trivially since $\tilde{Q} = Q$.  The illuminated state C$\rightarrow$D, with an elevated $\tilde{T} > T$, is isotropic with $\tilde{Q} \approx 0$ and the parameters are $\Pp =(1+2Q) >1$ and $\Pn = (1-Q) <1$.

A free elastomer has $\tilde{f} = 0$, in either the force-free state D, with temperature $\tilde{T}$, or A with temperature $T$. Changing from the reference temperature $T$ to $\tilde{T}$ and setting $\tilde{f} = 0$ in (\ref{nrubber}) then gives a contraction $\lambda\s{m} = (\Pn/\Pp)^{1/3}$ of its natural length along its director~\cite{warnerbook:07}.
Thus $\lambda\s{m} = [(1-Q)/(1+2Q)]^{1/3}$.
The large change between $\lambda = \lambda_m$ and $\lambda = 1$ is what makes LCEs promising working materials.
We henceforth describe optical response, thermal response being entirely analogous.
To realize a continuously operating engine, we subject the elastomer to a cyclical process through states A--B--C--D--A by changing the force and illumination.

It is easy to see from Fig.~\ref{fig2} that the net work delivered by the engine, per unit length of its LCE working material in its initial but unstretched state at $T$,  as it is taken around the cycle, is:
\bea
W &=& W\s{CD} - W\s{AB} \nonumber\\
 &=& \mu A_0\left(\int_{\lambda\s{m}}^{\lambda\s{h}} \tilde{f}(\lambda;\tilde{T})\d\lambda - \int_{1}^{\lambda\s{h}} \tilde{f}(\lambda;T) \d\lambda\right) \nonumber\\
&\equiv& \mu A_0\left(I_1 - I_2 \right) \label{work1}\\ &=& F(\lambda\s{h},\tilde{T}) - F(\lambda\s{h},T) + {\textstyle \frac{3}{2}}\mu A_0\left( 1 - (\Pp\Pn^2)^{1/3}\right) ,\nonumber
\eea
where the $\int \tilde{f}(\lambda;T)\d\lambda$ are reduced force integrals. $W$ is similar in form to the work done in the various gaseous $p$--$V$ cycles. We assume the moduli are comparable,
$\mu(T) \approx \mu(\tilde{T})$. Notice that the work $W$ delivered is greater than the difference established in the free energies per length of band in its unstretched state at C and B by a (positive) term ${\textstyle \frac{3}{2}}\mu A_0\left( 1 - (\Pp\Pn^2)^{1/3}\right)$ that arises during the cycle due to heat absorption.

One already sees from Fig.~\ref{fig2} that the pre-stretch, $\lambda\s{h}$, imposed before illumination, enhances work output considerably, just as a higher compression ratio improves conventional engines.  Further, taking a turbine allows one to extend the cycle to zero force (at a contraction equal to the natural illuminated length, $\lambda\s{m}$) which is impossible in two-wheel stretch engines~\cite{knezevic:13o}; see cycle a--B--C--s--a in Fig.~\ref{fig2}.  However for a soft solid, formidable problems exist in realising the above $W$.  These include sliding, and the consequential frictional losses, when finite tension differences in soft solids exist across the engine.  The remainder of this paper is concerned with overcoming these problems to get close to the above $W$ by using the turbine of Fig.~\ref{fig1}.  We will need to generalise the classical pulley result of Euler~\cite{euler:1762} to highly extensible belts.

The energy input per unit length of unstretched elastomer in the illumination process B--C is $\varepsilon n\s{dye} A_0$,
with $\varepsilon$ the appropriate photon energy and $n\s{dye}$ the number density of dye molecules. For our model of an isotropic elastomer,
the internal energy is a function of temperature only and is unchanging along an isotherm~\cite{treloar:05}.
Thus the heat input per unit length of unstretched elastomer during the isothermal process C--D equals the work
done $W\s{CD}$.
The efficiency is the ratio of the work done to the optical energy plus heat invested per cycle:
\be
\eta = \frac{W\s{CD} - W\s{AB}}{W\s{CD} + \varepsilon n\s{dye} A_0} = \left ( 1- \frac{I_2}{I_1} \right ) \! /\! \left ( 1+\frac{\varepsilon n\s{dye}}{\mu} \frac{1}{I_1} \right )
\label{eta1}
\ee
where $I_1$ and $I_2$ are the reduced force integrals in Eq.~(\ref{work1}) and depend on the order along the isotherms $T$ and $\tilde{T}$.  From Eq.~(\ref{nrubber}), they are clearly just the free energy changes per length of band in its unstretched state (reduced by $\mu A_0$) between the relevant $\lambda$s, and involve the $Q$ and $\tilde{Q}$ through the explicit expressions~(\ref{P-defs}) for the $P(Q)$s.

Since $I_1 > I_2$, the efficiency is always $0< \eta < 1$.
The ratio of material constants has been estimated as $\varepsilon n\s{dye}/\mu \approx 2$ ~\cite{knezevic:13o,knezevic:13a}.
Thus the efficiency is determined by the value of the order parameter $Q$ and the value of the pre-stretch $\lambda\s{h}$.
As an example we take a modest value for the dark-state order, $Q = 0.5$, which would yield $\lambda_m = 0.63$, or equivalently an elongation $1/\lambda_m = 1.58$ on recovering the nematic state from the isotropic, illuminated state.  In fact much greater elongations of $\sim 4$ are known.  With $Q = 0.5$ and $\lambda\s{h} = 3 \lambda\s{m}$, one obtains $\eta \approx 40 \%$.  We discuss dynamics and the influence on $\eta$ at the end.

We now follow the mechanical/thermal cycle in detail:
\nl
A closed band of photo-LCE is wound around the pair of spindles of Fig.~\ref{fig1}, the tops of which are rigidly coupled by
a loop of inextensible wire, thus ensuring they have equal angular velocities. The spindles have slightly tilted axes
which enables the elastomer to spontaneously follow a helical path down the spindles as they rotate. The horizontal cross-section of the spindles can be approximated by circles, the radii of which grow from the top to attain a largest value ($r$) at the middle section of the spindle, after which they then decrease.

Initially, a free elastomer belt in the nematic, reference state A at temperature $T$ comes on to the upper part of the spindles with $\lambda = 1$ and hence $f=0$.  Isothermally, it passes helically  from smaller to larger spindle radii, extending at each pass by a fraction equal to the ratio of the radii which material conservation requires as discussed later.
It suffers multiple-step extension during the process A--B in Fig.~\ref{fig2}.
State B, achieved when the belt enters the cylindrical section of the spindle with the largest radius $r$ (Fig.~\ref{fig1}), has the highest (pre-)stretch $\lambda = \lambda\s{h}$, and has force $f=f\s{B}$.
When in this region of radius $r$, the elastomer is illuminated and assumed to reach $\tilde{T}$ before it leaves the region; the extent of this region of the spindle is determined in part by dynamics since light absorption should proceed long enough to give sufficient dye conversion to attain the isotropic state before the band starts the contraction part of the cycle.  Along B--C the process is isometric ($\lambda=\lambda\s{h}$, an assumption of no slip along B--C).
At the point C, the force takes the highest value in the cycle $f\s{C}$, and the elastomer is in the isotropic state.
After passing the region of maximum radius $r$,
the elastomer spirals down the lower part of the spindles, C$\rightarrow$D. It moves now repeatedly from larger to smaller radii,
contracting at each pass by the ratio of the radii. It releases its elastic energy as work done on the spindles;
in the final state D the stretch reaches the minimum value $\lambda = \lambda\s{m}$, the natural length of elastomer at $\tilde{T}$. To remain with $\tilde{T}$ along C--D, because of \textit{cis} $\rightarrow$ \textit{trans} back reaction one must continue some level of illumination. Equally, to remain isothermal one needs external heat since entropy rises in a contracting elastomer. (In practice much of this heat input can come directly or indirectly from the optical source: Photons have energy in the range 2eV which is equivalent to $80\textrm{k}\s{B}T$ for $T \sim 300$K, that is, large energies per absorption are available from quantum efficiencies of less than 100\% in the \textit{trans} to \textit{cis} transition.  Equally, \textit{cis} to \textit{trans} back reactions yield energies per dye molecule that are large on the thermal energy scale and are spread over the network molecules.)
Finally, on removal of illumination, at zero force the elastomer gradually recovers and elongates from $\lambda\s{m}$ and $\tilde{T}$ to $\lambda=1$ and $T$ (D--A of the cycle).
The engine converts into mechanical work part of the optical energy elevating the elastomer to the effective temperature $\tilde{T}$.  Work is extracted by turning one spindle,
the left in Fig.~\ref{fig1}, against the external torque, $G\s{ext}$.

\section{Frictional losses} \label{friction}

\begin{figure}
  \includegraphics[width=8.7cm]{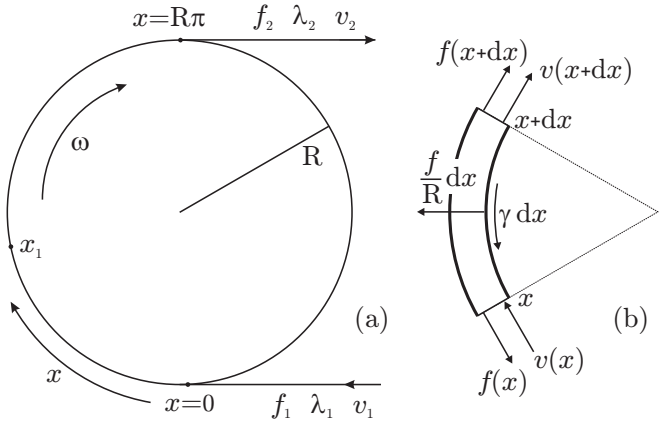}
  \caption{(a) A spindle with a belt around it with a tension varying with position $x$ around its circumference.
  (b) An expanded view of a short section of belt at $(x,x+\d x)$ with a radial force and a frictional
  shear force acting on the belt.
  }
\label{fig3}
\end{figure}

The useful work done per cycle is less than that of Eq.~(\ref{work1}) because of sliding friction. We analyze each pass of the elastomer around a spindle, considering differences in the tensions $f_1$ and $f_2$ of the incoming and outgoing sections; see Fig.~\ref{fig3}(a).

Classically~\cite{euler:1762}, there is an exponential increase of tension along
a belt passing around a wheel; see Fig.~\ref{fig3}(a) where tension increases from $f_1$ to $f_2$ over parts of $x=0$ to $x=\pi R$: Since the surface is curved, there is an inward force per unit length on the spindle from the belt of $f/R$; see Fig.~\ref{fig3}(b).  If the surface is rough, the normal force gives a frictional shear force (per unit length on the circumference) of $\gamma \propto f/R$.  The net shear force $\gamma \d x$ on an element $\d x$ balances the difference in tangential forces $f(x+\d x) - f(x) = (\d f/\d x) \d x$ across the element. Clearly $\frac{\d f}{\d x} = \gamma \propto f/R$ gives an exponential increase in $f(x)$.  There is an obvious limitation to the classical analysis that is important in soft solids where substantial stretch $\lambda$ must accompany increasing $f$:  The belt has to slide on the spindle to extend so that $f$ increases with $x$.  It cannot translate with the circumferential speed $v = \omega R$ for all its contact length with the spindle, but rather travel faster.  Changing strain implies changing stored
elastic energy. Sliding implies frictional losses during the transmission of power.

Fig.~\ref{fig3}(a) assigns a $\lambda(x)$ corresponding to a $f(x)$. In particular $\lambda = \lambda_1$ for the belt incoming on to the spindle, and $\lambda = \lambda_2$ when it emerges with $f_2$.
Take temperature constant along the belt in contact with the spindle; increasing tension $f_2>f_1$ then implies $\lambda_2>\lambda_1$.
Soft solids are essentially incompressible under extension, hence material conservation dictates volume conservation.
The sectional area $A= A_0/\lambda$ must diminish as the belt extends by $\lambda$. The volume flux of belt on to the spindle is
$v_1A = v_1A_0/\lambda_1$ and must be matched by $v(x)A_0/\lambda(x)$ at a general point. Dividing through by $A_0$ gives
\be
v_1/\lambda_1 = v(x)/\lambda(x) = v_2/\lambda_2,
\label{speed}
\ee
which is the length of band per unit time passing, were it to be in its unstretched state.  This flux of length of band is equivalent to $\omega R /\lambda_1$ when the band is not sliding.  Since the final stretch is greater than the initial, $\lambda_2>\lambda_1$, it follows $v_2 > v_1$ --- the belt starts going faster than the spindle.  There is \textit{sliding} friction.
Neglecting inertia \cite{bechtel:99}, we get:
\be
\frac{\d f}{\d x} = \mu_{\rm k} \frac{f}{R},
\label{fplength}
\ee
where $\mu_{\rm k}$ is the coefficient of kinetic (sliding) friction giving $\gamma = \mu_{\rm k} f/R$. Since $f_2>f_1$, friction
inhibits sliding in the $+x$ direction. Solving (\ref{fplength}) gives
\be
f(x) = f_2 \exp \left [ \frac{\mu_{\rm k}}{R} (x - \pi R) \right ],
\label{tension}
\ee
which differs from the classical result in its use of kinetic friction coefficient, $\mu\s{k}$, and the passage of energy to
friction and to elastic potential.
The tension varies from $f=f_2$ at $x=\pi R$ down to $f=f_1$ at an $x=x_1$ given by
\be
x_1 = R \left [ \pi - \frac{1}{\mu_{\rm k}}\ln \left ( \frac{f_2}{f_1} \right ) \right ].
\label{x1}
\ee
In the initial section $(0,x_1)$, the tension retains its incoming value $f=f_1$.
It is gradients $\d f/\d x$ that transfer force (torque) to the spindle in the region $(x_1,\pi R)$.
There is a region $(0,x_1)$ without torque if $x_1>0$, that is if $f_2<f_1 \rm{e}^{\mu_{\rm k}\pi}$.

The power delivered to the spindle, in one step of A--B of Fig.~\ref{fig2} with $f_2>f_1$ and the spindle turning clockwise (see Fig.~\ref{fig3}(a)), is the speed of its surface, $v_1$, times the force exerted on it:
\be
P\s{w} = v_1\!\! \int_{x_1}^{\pi R}\!\!\!\! \d x \mu\s{k} \frac{f(x)}{R}  = v_1\!\!  \int_{x_1}^{\pi R}\!\!\!\! \d x \frac{\d f}{\d x} = v_1 (f_2 - f_1).
\label{pw}
\ee
The term $-v_1 f_1$ is the power given by the spindle to the incoming band. The power $v_1 f_2$ is the portion of
the power $v_2 f_2$ delivered by the more tense band that actually finds its way to the spindle. Thus, all the useful power
is delivered via the region that is slipping.
When the band slips as soon as it makes contact with the wheel, slip is complete and $x_1 = 0$ in (\ref{x1}); then $f_2 = f_1 \rm{e}^{\mu\s{k} \pi}$ takes its
maximal value, as does $P\s{w}$ in Eq.~(\ref{pw}); $P\s{w} = v_1 f_1 (\rm{e}^{\mu\s{k} \pi} - 1)$.

The power lost to friction in a step of A--B is from the forces $(\d f/\d x)\d x$ acting on elements $\d x$ moving at speeds $v(x) - v_1$ relative to the elements:
\be
P\s{f} = \int_{x_1}^{\pi R} \d x \frac{\d f}{\d x} [ v(x) - v_1 ],
\label{pf}
\ee
with $x_1$ given by (\ref{x1}). The $v_1$ part is trivially $-v_1 (f_2-f_1)$, Eq.~(\ref{pw}). The first term integrates by parts to
$(f_2 v_2 - f_1 v_1) - \int_{x_1}^{\pi R} \d x f(x) \d v(x)/\d x$. The latter part has
$\d v/\d x = (v_1/\lambda_1) (\d\lambda/\d x)$, and becomes:
\bea
- \frac{v_1}{\lambda_1} \int_{x_1}^{\pi R} \d x &f(x)& \frac{\d\lambda}{\d x} =
- \frac{v_1}{\lambda_1} \int_{\lambda_1}^{\lambda_2} \d\lambda f(\lambda) \nonumber \\
&=& - \frac{v_1}{\lambda_1} \left [ F(\lambda_2,T) - F(\lambda_1,T) \right ],
\label{elastic}
\eea
where we used Eq.~(\ref{nrubber}) with $\Pp = \Pn =1$, since the elastomer is at $T$.
Overall, the power lost to friction is:
\be
P\s{f} = f_2(v_2 - v_1) - \frac{v_1}{\lambda_1} [ F(\lambda_2,T) - F(\lambda_1,T) ],
\label{pff}
\ee
and does not involve the friction coefficient $\mu\s{k}$, unless slippage is complete ($x_1 =0$). In the process A--B of the engine,
$\lambda_2>\lambda_1$ is realized by moving the elastomer belt in each pass from a spindle surface of
a smaller radius $r_1$ to a surface of a larger radius $r_2$. The velocities are $v_1 = \omega r_1$ and $v_2 = \omega r_2$.
Then using (\ref{speed}) one gets $\lambda_2/\lambda_1 = r_2/r_1>1$, as required.

In process C--D, in each turn around a spindle the isotropic elastomer with $\tilde{T}$ goes from a larger to a smaller radius, $r_1>r_2$ leading to $v_1 > v_2$ and $\lambda_1 > \lambda_2$ since $\lambda_2/\lambda_1 = r_2/r_1<1$.  Hence the band goes from a higher to lower force, $f_1>f_2$; see Fig.~\ref{fig3}(a).
Note that the spindle's direction of turn is fixed in the clockwise direction by the engine operation as a whole,
independently of the relative size of the forces.
As before, there is an $x_1$ with no slip for $x <x_1$, while slippage occurs for $x$ in the range $(x_1,\pi R)$.
Repeating the above, the power delivered and the frictional power lost are:
\bea
P\s{w} &=& v_1 (f_2 - f_1) < 0\, , \\
\label{isopw}
P\s{f} &=& f_2(v_2 - v_1) + \frac{v_1}{\lambda_1} [ F(\lambda_1,\tilde{T}) - F(\lambda_2,\tilde{T}) ].
\label{isopf}
\eea

It remains to analyze the elastomer belt going around surfaces of radius $r$ in the process B--C.
Now the elastomer at force $f\s{B}$ and temperature $T$ comes on the spindle where the illumination begins; see Fig.~\ref{fig1}.
In Fig.~\ref{fig3}(a) we now have $f_1 = f\s{B}$ and $\lambda_1 = \lambda\s{h}$. We assume that the change
$T$ to $\tilde{T}$ occurs without slippage and is complete before the elastomer reaches the point C in Fig.~\ref{fig1}, possibly after several passes around the central section.  (Force rising from $f\s{B}$ to $f\s{C}$ without slippage assumes $\d f /\d x < \mu\s{s} f/r$ where $\mu\s{s}$ is the static friction coefficient.  Elongation is fixed at $\lambda = \lambda\s{h}$ and the variation of $f$ with $x$ is from the changing $\Pn$ and $\Pp$ factors in expression~(\ref{nrubber}) for $f$.)  After $x_1$, slippage then occurs and the force drops from $f\s{C}$ to $f\s{S}$ (in Fig.~\ref{fig3}(a) $f_2=f\s{S}$, $\lambda_2= \lambda\s{S}$). The state S is marked in Figs.~\ref{fig1} and~\ref{fig2}. The net power delivered to the spindles on transit of the cylindrical section with radius $r$ is
\be
P\s{w} = \omega r (f\s{S} - f\s{B}).
\label{pwcs}
\ee
For the frictional losses, expression (\ref{isopf}) with the appropriate values of stretches, velocities and forces is still applicable since process C--S is part of C--D.

Excluding the middle spindle section of radius $r$, the power delivered to the spindle in each extension step of A--B is given by expressions of the form (\ref{pw}) and is clearly positive
as one expects if $f_2 > f_1$. However, the total power delivered in A--B is actually negative (see Fig.~\ref{fig2}) since the band emerges stretched from A--B.
This is due to the fact that one must add the highly negative contribution $-\omega r f\s{B}$ of (\ref{pwcs}) to the set of positive contributions of the form (\ref{pw}).
Similarly, in C--D the highly positive term $\omega r f\s{S}$ of (\ref{pwcs}) is added to negative contributions (\ref{isopw}), making the total
power delivered in C--D positive since the elastomer contracts.

We now calculate the total power $P_{nk}$ delivered to the spindles by the elastomer for a $n$--stage contraction process C--D and
$k$--stage extension process A--B.
When the engine runs at a constant velocity $\omega$, the net torque acting on each of the spindles is zero. For simplicity we
shall neglect frictional forces at the bearings. Then one can express the balance of torques on the each spindle separately.
Beside the torques produced by the elastomer, one should take into account the torques due to the inextensible wire which
couples the spindles, and the external torque $G\s{ext}$ acting on the left spindle (see Fig.~\ref{fig1}). In such a way,
one obtains a pair of equations which enable one to determine $G\s{ext}$, and thus the useful power $P_{nk} = \omega G\s{ext}$:
\be
P_{nk} = \omega  \sum_{i=1}^{n} f_{i+1} (r_i - r_{i+1}) - \omega \sum_{j=1}^{k} f_{j+1}' (r_{j+1}' - r_j').
\label{powernk}
\ee
Here, $f_{i+1}$ and $f_{j+1}'$ are the forces in the $i$--th step of C--D and in the $j$--th step of A--B process, respectively, while
$r_i$ and $r_j'$ are the corresponding spindle radii.
In particular, $f_2 \equiv f\s{S}$ and $r_1 \equiv r$ (force $f_1$ would be $f_1 \equiv f\s{C}$).
Clearly, since $r_i > r_{i+1}$ for all $i$, the first sum on the right hand side of (\ref{powernk}),
$P_n = \omega  \sum_{i=1}^{n} f_{i+1} (r_i - r_{i+1})$, is positive.
Similarly, we have $f_{k+1}' \equiv f\s{B}$ and $r_{k+1}' \equiv r$ (force $f_1'$ would be zero). On the other hand,
$P_k = - \omega \sum_{j=1}^{k} f_{j+1}' (r_{j+1}' - r_j')$ is negative since $r_{j+1}'>r_j'$.

Using relations (\ref{pff}) and (\ref{isopf}), the useful power is
\bea
P_{nk} &=& \frac{\omega r}{\lambda\s{h}} \sum_{i=1}^{n} [ F(\lambda_i, \tilde{T}) - F(\lambda_{i+1},\tilde{T})] - \sum_{i=1}^{n} P_{{\rm f}, i}
\nonumber \\
&-& \frac{\omega r}{\lambda\s{h}} \sum_{j=1}^{k} [ F(\lambda_{j+1}', T) - F(\lambda_{j}',T)] - \sum_{j=1}^{k} P'_{{\rm f}, j} \nonumber \\
&=& \frac{\omega r}{\lambda\s{h}} W - \sum_{i=1}^{n} P_{{\rm f}, i} - \sum_{j=1}^{k} P'_{{\rm f}, j},
\label{areapowernk}
\eea
where $\lambda_i$ and $\lambda'_j$ are the stretches, and $P_{{\rm f}, i}$ and $P'_{{\rm f}, j}$ are the frictional power losses in the
contraction and extension steps respectively.
We now sketch a geometrical interpretation of (\ref{areapowernk}).
The first term of the last line is the maximum useful mechanical power $P = W \omega r/\lambda\s{h}$,
with $\omega r/\lambda\s{h}$ being the flux of unstretched length (\ref{speed}).
The maximum mechanical work $W$ done per unit length of unstretched elastomer in one cycle is given by the area in the solid line A--B--C--D--A in
Fig.~\ref{fig2}.
The useful work done per unit length of unstretched elastomer in one cycle $W_{nk} = P_{nk}/(\omega r/\lambda\s{h})$
is the gray shaded area enclosed in Fig.~\ref{fig2} (where for simplicity the change in radii in each contraction and extension step
is taken to be the same; see comments below).
Thus, the work done in one cycle to overcome friction during the contraction process is equal to the area of the white regions between the solid curve C--D and the gray shaded area (corresponding frictional power losses are $\sum_{i=1}^{n} P_{{\rm f}, i}$). Similarly, the area of regions between the gray shaded area and the solid curve A--B corresponds to the term $\sum_{j=1}^{k} P'_{{\rm f}, j}$ of (\ref{areapowernk}).

By contrast, the two-wheel cycle a--B--C--s--a dashed in Fig.~\ref{fig2} has larger frictional losses: the areas between the isotherms a--B and C--s and the horizontal dashed lines respectively above and below them are much larger relative to the enclosed area a--B--C--s--a (the maximum theoretical work).  The inaccessibility of an $\tilde{f} = 0$ state means the enclosed area is smaller than the turbine's, even though the same optical energy input to go from B to C is required, further underscoring the superior efficiency of a turbine.

Notice that the more steps $n$ and $k$, the more useful work per cycle is done. We shall compare
the power $P$ corresponding to an infinite number of extension and contraction steps, with the power $P_{nk}$
given by expression~(\ref{powernk}), and determine how many steps $n$ and $k$ one needs to achieve, say,  $P_{nk}/P=0.9$.

As an illustration, we take the change of radii in the contraction process C--D to be the same for each step and equal to $\Delta r = (r-r_{n+1})/n$,
where $r_{n+1}$ is the final radius in the contraction process. Similarly, for the extension process A--B we take
$\Delta r' = (r-r'_1)/k$, where $r'_1$ is the initial radius in the extension process.
From relations (\ref{speed}) it follows:
$r/\lambda\s{h} = r_{i}/\lambda_{i} = r_{n+1}/\lambda\s{m}$ and
$r'_1/1 = r'_{j}/\lambda_{j} = r/\lambda\s{h}$. Here we used the facts that the stretch that corresponds to the radius
$r_{n+1}$ is actually the natural stretch $\lambda\s{m} = (\Pn/\Pp)^{1/3}$, and that the stretch corresponding to radius $r'_1$
is $\lambda=1$. The stretch steps in contraction and extension processes are then $\Delta \lambda = \Delta r \lambda\s{h}/r
= (\lambda\s{h} - \lambda\s{m})/n$ and $\Delta \lambda' = \Delta r' \lambda\s{h}/r = (\lambda\s{h} - 1)/k$, respectively.
The choice $\Delta \lambda = \Delta \lambda'$ was made in Fig.~\ref{fig2}, and cones taken in Fig.~\ref{fig1}, but these restrictions are not necessary.

The powers $P_{nk}$ and $P$ can be expressed as
\bea
P_{nk} \!&=&\! \mu A_0 \frac{\omega r}{\lambda\s{h}} \!\left[
\Delta \lambda \sum_{i=1}^{n} \left( \Pp (\lambda\s{h} - i \Delta \lambda) - \frac{\Pn}{(\lambda\s{h} - i \Delta \lambda)^2} \right)\right. \nonumber \\
&-& \!\left. \Delta \lambda' \sum_{j=1}^{k} \left( 1 + j \Delta \lambda' - \frac{1}{ (1 + j \Delta \lambda')^2} \right) \right], \\
P \!&=& \!\mu A_0 \frac{\omega r}{\lambda\s{h}}\! \left[ \int_{\lambda\s{m}}^{\lambda\s{h}} \d\lambda (\Pp \lambda - \frac{\Pn}{\lambda^2}) \nonumber
- \int_1^{\lambda\s{h}} \d\lambda (\lambda - \frac{1}{\lambda^2}) \right].
\label{PPinf}
\eea
Taking $Q=0.5$ and $\lambda\s{h} = 3 \lambda\s{m}$, one finds $P_{nk}/P =0.9$ for $n=15$, $k=15$.

The maximum mechanical work done per unit length of unstretched elastomer in one cycle $W = P/(\omega r/\lambda\s{h})$
is given in Eq.~(\ref{work1}). The simple integrals above at constant temperatures return us to the free energies in
Eq.~(\ref{nrubber}).

\section{Conclusions} \label{concs}

In summary, the optical contraction of photo-LCEs can be used to harness optical energy to generate mechanical energy.
Our mechanical turbine-based engine utilizes more effectively the optothermal cycle than the two-wheel engine~\cite{knezevic:13o}.
Soft, extensible photo-solids deliver large amounts of work, but their extensions and contractions lead to sliding and hence concomitant
frictional losses. We analyzed such losses and calculated the fraction of work lost due to them.

That our proposed photo-conversion method has moving parts gives it a disadvantage over conventional photo-voltaics, though rubber is highly durable and tough --- for instance car tyres survive long use in harsh, abrasive conditions.  Another difficulty, that could perhaps be solved by chemical design, is degradation from the effect of UV light.  The role of dynamics when attempting to attain the above efficiencies also remains a question.  As we noted, the residence time in the isometric state B--C needs to be long enough that light penetrates deeply to convert the whole thickness of the band to the isotropic state. For intense beams this requires a non-linear bleaching wave (of light-induced dye conversion) to traverse the thickness \cite{knezevic:13} just as that part of the band leaves the section of the spindle with maximal radius $r$.  Longer residence simply gives time for back reaction without mechanical work being delivered.  Light energy is wasted and the efficiency drops from the values emerging from the above analysis. Thus light intensity, band thickness, residence in the isometric state, and rotational rate are all related aspects of the dynamics and need optimisation by future studies.


M. K. thanks the Winton Programme for the Physics of Sustainability and the Cambridge International Trust for financial support.  M. W. thanks Professor Peter Palffy-Muhoray for introducing him to these studies and for advice over many years.

\end{document}